\documentclass{kluwer}    

\usepackage[dvips]{epsfig}
\newdisplay{guess}{Conjecture}

\begin{document}                                                                                   
\begin{article}
\begin{opening}         
\title{The Submillimeter Extragalactic Background and its implication for
the star formation history of the Universe} 
\author{Guilaine \surname{Lagache}, Jean-Loup \surname{Puget} and Richard \surname{Gispert$^{\dag}$}}  
\runningauthor{Lagache, Puget \& Gispert}
\runningtitle{The Submm Extragalactic Background. Star formation history of the Universe}
\institute{IAS, b\^at 121, Universit\'e de Paris XI, 91405 ORSAY Cedex}
\date{August 13, 1999}
\begin{abstract}
The recently discovered Submillimeter Extragalactic Background
(submm EB) has opened new perspectives on our understanding of galaxy
evolution. We detail in this paper one of the major cosmological 
consequences of this discovery, the submm luminosity density 
history of the Universe.
\end{abstract}
\end{opening}           

\vspace{-.3cm}
\section{The Extragalactic Background (EB)}
Our knowledge of the early epochs of galaxies has recently
increased thanks to the observational evidence provided by 
UV/Vis/Near-IR, far-IR and submm surveys 
of high-redshift objects. In a consistent picture,
the epoch of galaxy formation can also be observed by its imprint
on the background radiation which is produced by the line-of-sight
accumulation of all extragalactic sources. \\
The search for the UV, optical, near-IR and mid-IR EB
(from 2000~$\AA$ to 15 $\mu$m) 
obtained by summing up the contributions of galaxies
using number counts
currently gives only lower limits. However, in the optical and near-IR
the flatening of the faint counts obtained in the Hubble Deep Field
suggests that we are now close to convergence.
Thus we already have a good determination of the background in the optical and near-IR.\\
At longer wavelengths ($\lambda>$100 $\mu$m), a submm EB was predicted to exist 
as early as the 1970s. It has been detected for the first time only in 1996
by Puget et al. thanks to COBE data. Its detection, 
now confirmed by several other studies
(Fixsen et al. 1998; Hauser et al. 1998; Lagache et al. 1999)
has provided new perspectives for understanding galaxy evolution.
The amount of energy in the submm EB is rather large.
It exceeds the predictions based on extrapolations of the starburst
galaxies as seen in the IRAS deep surveys by about a factor of 3
(Franceschini et al. 1994).
Thus, it requires other sources, for example, spheroidal systems radiating
mostly in the far-IR during their early evolution, or a high
formation rate of massive stars at the early stages in the 
evolution of elliptical galaxies. Whatever the nature of the
emitting sources, the background implies a very strong cosmological
evolution from the local galaxies to the more distant ones
(Guiderdoni et al. 1998; Franceschini et al. 1998),
with an enhancement of the IR emission at early stages of 
the galaxy evolution. This very strong evolution is 
also supported by the
comparaison between the energy contained in the optical and 
submm EB. The sumtotal of the fluxes of the EB from
6 to 1000 $\mu$m is about 3.3 10$^{-8}$ W m$^{-2}$ sr$^{-1}$ compared to
2 10$^{-8}$ W m$^{-2}$ sr$^{-1}$ between the UV band
and 6 $\mu$m. For the local galaxies, the energy ratio between IR
and optical wavelengths is much smaller
than that measured from the background (about 0.4 compared to 1.6).
This simply reflects a strong change of galaxy properties
between the local Universe and the more distant Universe.
This change must affect the Star Formation (SF) history 
derived only from optical and near-IR observations of high-redshift
objects. A significant fraction of SF might be
hidden in heavily extinguished galaxies. These galaxies 
can be totally missed by optical surveys.

\vspace{-.3cm}
\section{Star formation history}
The history of the cosmic Star Formation Rate (SFR) can be derived
from deep optical surveys assuming that (1) the stellar Initial Mass
function is universal, (2) the far-UV light is proportionnal
to the SFR and (3) extinction is negligible. The presence of dust
which absorbes the UV starlight makes this last assumption not valid.
The corrections needed to account for the extinction
are rather uncertain and there is much controversy about
the value of this correction. Moreover, the SFR deduced from
optical surveys can be underestimated if there is a significant
population of objects so obscured that they are not detected
in these surveys. A way to avoid this dust contamination
is to derive the SFR from the IR/submm surveys. However, so far, the catalogues
of faint submm sources with reliable redshifts are not large 
enough to reconstruct the history of the SFR.
But strong constraints can be provided by the inversion
of the submm EB.
\subsection{Inversion of the submm EB}
For a Universe following the Robertson-Walker metric, the intensity of
the submm EB ($I_{\nu}$) can be simply related to the
number of sources per Mpc$^3$ ($N_{z}$) having a luminosity
$L_{\nu_0}$ using the expression:
\begin{equation}
\label{eq_principale}
\nu I_{\nu}= \frac{c}{4 \pi} \int_{z=0}^{\infty} \int_{\nu_0=\nu}^{\infty} N_{z} L_{\nu_0} \frac{dt}{dz} d\nu_0 \quad \delta(\nu_0 - \frac{\nu}{1+z}) dz
\end{equation}
Choosing a spectral energy distribution $L_{\nu_0}$ and a cosmological model
(i.e dt/dz), the inversion of Eq. \ref{eq_principale}
will give $N_{z}$. The variation of the luminosity density 
as a function of redshift $\varphi(z)$, is then derived as follows:
\begin{equation}
\label{GPL_k}
\varphi(z)= N_{z} \int_{\nu_0}  L_{\nu_0} d\nu_0 \quad L_{\odot}/Mpc^3
\end{equation}
Using such a formalism, the production rate of the far-infrared
radiation as a function of redshift can be deduced using only one
assumption: the spectrum of the galaxy $L_{\nu_0}$. 
Since the submm EB seems to be dominated
by luminous IR galaxies, as indicated by the ISO 
(Aussel, 1998) and SCUBA (e.g. Lilly et al. 1999) results, 
we choose a spectral energy distribution $L_{\nu_0}$ typical of
starburst galaxies (Maffei, 1994; Guiderdoni et al. 1998).
To determine $\varphi(z)$ (Eq. \ref{GPL_k}), 
we choose to explore the cases given by the combinations of:
(1) three infrared galaxy luminosities: 3 10$^{12}$ L$_{\odot}$, 5 10$^{11}$ L$_{\odot}$ and 
3 10$^{10}$ L$_{\odot}$ 
(2) two values for the dust spectral index (-1.7 and -2)
and (3) three cosmological
models defined by the set of parameters h, $\Omega_0$ and $\Omega_{\Lambda}$
(h=0.65, $\Omega_0$=0.3, $\Omega_{\Lambda}$=0.7; h=0.65, $\Omega_0$=0.3, $\Omega_{\Lambda}$=0, 
and h=0.65 $\Omega_0$=1, $\Omega_{\Lambda}$=0) which fixes the dt/dz.
The basic algorithm for finding N$_z$ is based on Monte Carlo simulations.
N$_z$ is sampled at a few redshift values and linearly interpolated between these values
for computing the term on the right of Eq. \ref{eq_principale}. It has been assumed that beyond z=13,
N$_z$=0. The best evaluation (minimum $\chi^2$) has been obtained 
by exploring a wide range of randomly distributed
values of  N$_z$. Error bars are estimated by keeping the computed 
submm EB (1) within the submm EB uncertainties at each sampled frequency and (2) greater than 
the lower submm EB limit at 850 $\mu$m from Barger et al. (1999).
By using an iterative method for progressively reducing the range of explored values, 
we reach convergence for each case studied with a reasonable number of 
hits ($\sim$ 50000). For the submm EB ($I_{\nu}$ in Eq. \ref{eq_principale}), 
we use the FIRAS and the DIRBE 140 and 240 $\mu$m determinations of Lagache et al. (1999).
$\varphi(z)$ at z=0 has been fixed following Soifer \& Neugeubauer (1991):
for h=0.65, we have $\varphi(z=0)$=1.07 10$^8$ L$_{\odot}$/Mpc$^3$.
\subsection{Main result}
Results for $\varphi(z)$ are presented in Fig. 1a for a fixed
cosmological model and three different luminosities. We note that
the combination of all parameter sets gives results consistent within a factor of two.
As was expected, constraints on $\varphi(z)$ are very weak below
redshift 1 (no EB values below 140 $\mu$m) and above 4 (very low signal
to noise ratio of the FIRAS submm EB above 800 $\mu$m). Between redshifts
1.5 and 3.5, the submm EB gives strong constraints. The luminosity density
is about 10 times higher at z=1.4 than at z=0 and it 
is nearly constant up to redshift 4. Data points derived from optical observations
(Fig. 1b)\footnote{To compare the star formation rate with the luminosity density, 
we use: $\frac{SFR}{M_{\odot} yr^{-1}}= \frac{L_{IR}}{7.7 10^9 \quad L_{\odot}}$
(Guiderdoni et al. 1998)} are lower by a factor of about 5 than our determination. 
This clearly shows
that the optically-derived SFR is severely underestimated due to the presence
of dust.

\begin{figure*}
\begin{minipage}{7.cm}
\epsfxsize=6.cm
\epsfysize=5.cm
\epsfbox{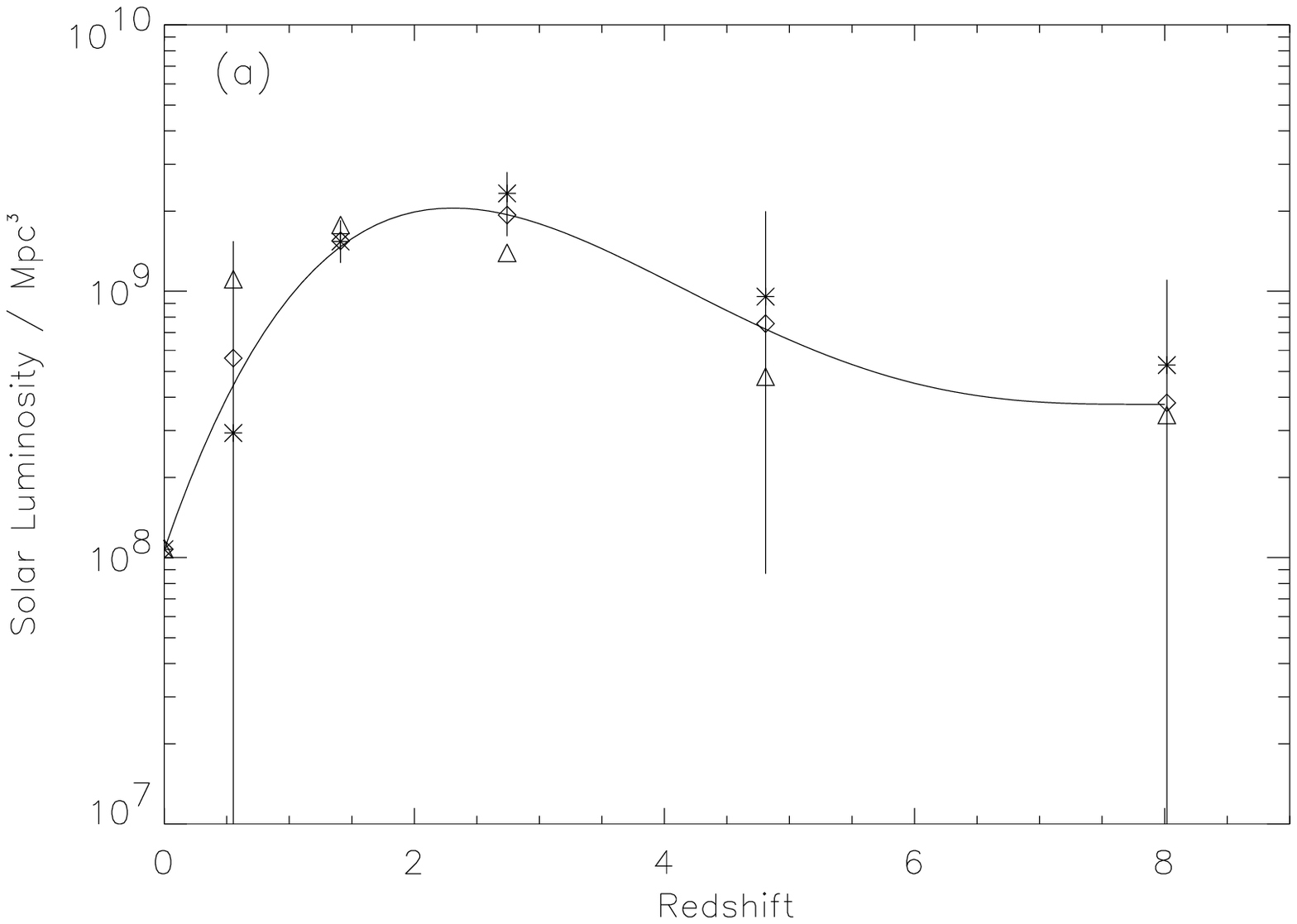}
\end{minipage}
\hspace{-3.cm}
\begin{minipage}{7.cm}
\epsfxsize=6.cm
\epsfysize=5.cm
\hspace{1.5cm}
\epsfbox{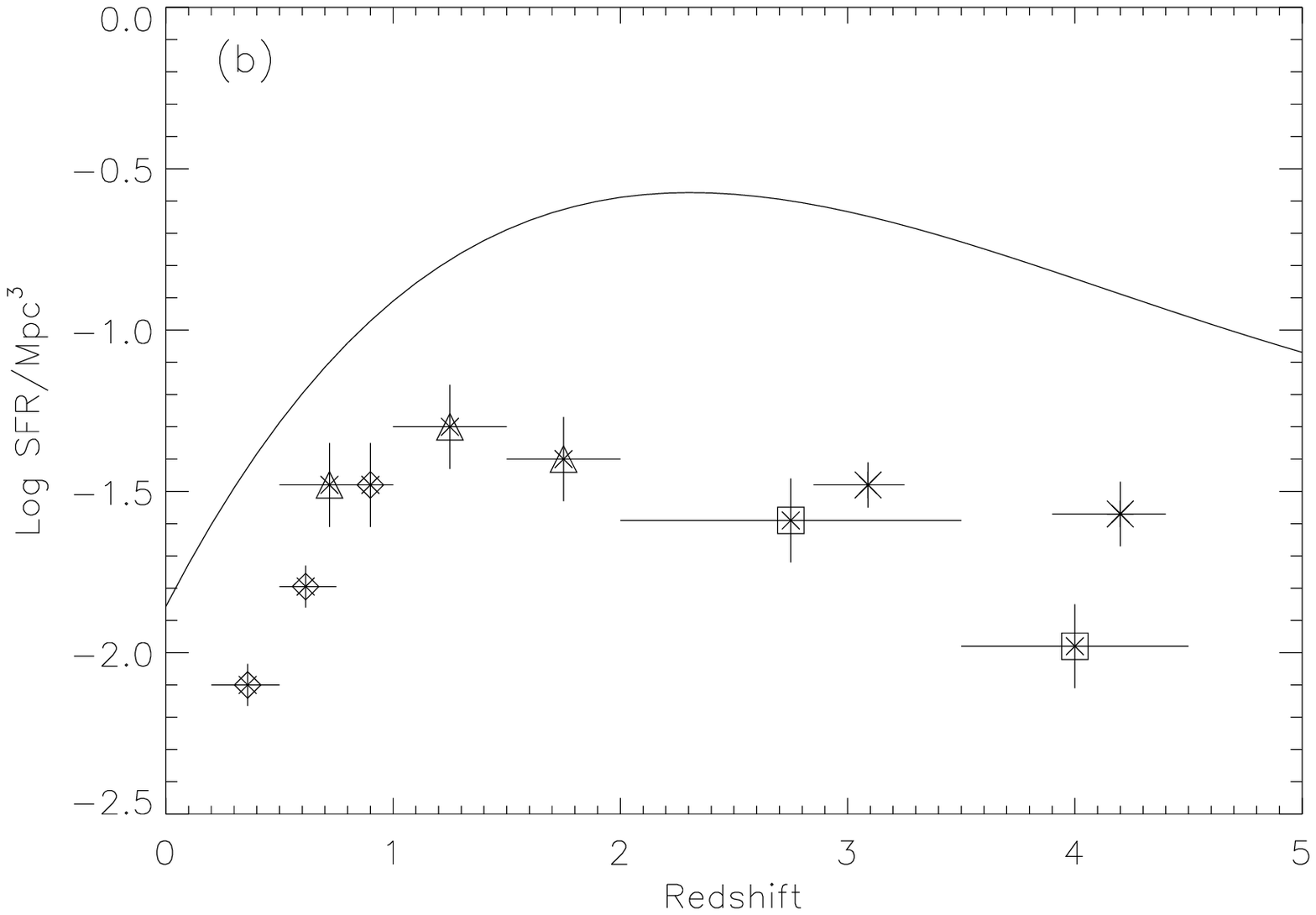}
\end{minipage}\\
\vspace{-.2cm}
\caption{\label{fig_sfr} (a) Luminosity density distribution as derived from the submm EB
for a given cosmological model (h=0.65, $\Omega_0$=0.3, $\Omega_{\Lambda}$=0.7) and
dust spectral index ($\alpha=2$) with three
different luminosities (stars: 3~10$^{12}$ L$_{\odot}$, diamonds: 5~10$^{11}$ L$_{\odot}$, 
triangles: 3~10$^{10}$ L$_{\odot}$). The error bars represent the uncertainties 
derived for the star points.
Also represented is the analytical function that reproduces the
case L=5~10$^{11}$ L$_{\odot}$, $\Omega_0$=1 and $\Omega_{\Lambda}$=0. The analytical function
is 10$^{az^3 + bz^2 + cz + d}$ with a=-2.42 10$^{-3}$, b=5.78 10$^{-2}$, c=1.36 and
d=8.03. (b) SFR
derived from the submm EB (analytical function, continuous line) and UV/Vis/Near-IR
observations (diamonds: Lilly et al. 1996; triangles: Conolly et al. 1997;
squares: Madau et al. 1996 and crosses: Steidel et al. 1999)}
\end{figure*}

\theendnotes

\end{article}
\end{document}